\documentclass[showpacs,aps,amsmath,amssymb,twocolumn]{revtex4-1}
 
\usepackage[]{graphicx}
\usepackage{color}
\usepackage{bbold}
\usepackage{braket}
\usepackage[colorlinks=true, pdfstartview=FitV, linkcolor=blue, citecolor=red, urlcolor=black]{hyperref}
\newcommand{\be}{\begin{equation}}
\newcommand{\ee}{\end{equation}}
\newcommand{\ben}{\begin{eqnarray}}
\newcommand{\een}{\end{eqnarray}}
\newcommand{\bes}{\begin{subequations}}
\newcommand{\ees}{\end{subequations}}
\def\bal#1\eal{\begin{align}#1\end{align}}

\newcommand{\bfi}{\begin{figure}}
\newcommand{\efi}{\end{figure}}
\newcommand{\bc}{\begin{center}}
\newcommand{\ec}{\end{center}}
\newcommand{\sech}{\mbox{sech}}

\begin{document}
\title{New Results For Cuscuton Multi-Field Brane}
\author{D. Bazeia$^{1}$ and A. S. Lob\~ao Jr.$^{2}$}
\affiliation{$^1$Departamento de F\'\i sica, Universidade Federal da Para\'\i ba, 58051-970 Jo\~ao Pessoa, PB, Brazil\\
$^2$Escola T\'ecnica de Sa\'ude de Cajazeiras, Universidade Federal de Campina Grande, 58900-000 Cajazeiras, PB, Brazil}
\begin{abstract}
We investigate braneworld models in the presence of the several scalar fields with generalized dynamics. A new mechanism to control the internal structure of  brane in modified gravity is used to induce  changes in the models in the presence of cuscuton dynamics. Examples with two- and three-field are considered, and we have found that the inclusion of the cuscuton can significantly alter the profile of the brane, causing unique changes in the warp factor, energy density, stability potential and the corresponding zero mode.
\end{abstract}

\maketitle
{\it {1. Introduction.}} Thick brane models are now standard generalizations of the pioneer Randall-Sundrum model \cite{Randall:1999vf}, and they are supported by the inclusion of scalar fields in the corresponding Einstein-Hilbert action \cite{Goldberger:1999uk, Skenderis:1999mm, DeWolfe:1999cp, Csaki:2000fc,Brito:2001hd,Campos:2001pr,Kobayashi:2001jd,Bazeia:2004dh}. In this representation, the scalar fields induce an internal structure in the brane solutions, enabling the emergence of new behaviors that are not verified in the thin braneworld scenario. Since the first work on thick brane in the late 1990s, many other proposals for generalized brane models have been investigated, mainly by the inclusion of new invariant terms in the action. Many of the ideas in this regard involve the inclusion of generalized dynamics known as $K$-fields \cite{Koley:2005nv, Pal:2007gp, Bazeia:2008zx, Liu:2009ega, Bazeia:2013euc,Zhong:2012nt}, $F(R)$-brane \cite{Afonso:2007gc, Dzhunushaliev:2009dt, Zhong:2010ae, Liu:2011wi}, teleparallel gravity \cite{Yang:2012hu,Menezes:2014bta, Guo:2015qbt}, models with quadratic gravity terms, such as Gauss-Bonnet \cite{Charmousis:2002rc,Kofinas:2003rz,Cho:2001nf,Maeda:2003vq}, and other possibilities.

An interesting consequence that usually arises in brane models with non-canonical action is the modification of the internal structure of the solutions. In this perspective, we have investigated that the inclusion in the action of non-standard dynamics for the source fields of the model, such as Born-Infeld or cuscuton, can significantly modify the internal structure of important quantities such as energy density, warp factor and zero mode. Cuscuton-like dynamics have been used to investigate various scenarios of generalized gravity \cite{Afshordi:2007yx,Afshordi:2006ad}, leading to a better understanding of how gravity works. Studies on braneworld models with cuscuton-like dynamics have been produced recently, some of these results can be verified in \cite{Bazeia:2012br,Ito:2019fie,Iyonaga:2018vnu,Andrade:2018afh,Bazeia:2021jok,Bazeia:2021bwg,Rosa:2021myu}. More recently, in the work \cite{Kohri:2022vst} the authors discussed cosmology based on cuscuton gravity to suggest understanding the anomaly of the observational Helium abundance reported by the EMPRESS collaboration \cite{Matsumoto:2022tlr}.

In the recent investigation \cite{Bazeia:2022BL}, we followed the proposal presented in \cite{Bazeia:2006ef,Ahmed:2012nh,deSouzaDutra:2014ddw} to show that it is possible to control the internal structure of the solutions obtained in brane models supported by several scalar fields. We uncovered that when the centers of the field solutions move away from each other, one changes the profile of the brane, in particular the warp factor, the energy density and the zero mode of the potential which describes linear stability of the geometry of the brane. In this sense, in the present study we are motivated to understand the new effects that should arise in the internal structure of brane solutions when we introduce non-canonical dynamics for the source fields. In particular, motivated by the Bloch brane already investigated in Ref. \cite{Bazeia:2004dh} and by the mechanism recently described in Ref. \cite{Bazeia:2022BL} to control the internal structure of the brane, we are interested in investigating cuscuton-like dynamics in the multi-field scenario, that is, in the presence of several scalar fields, because this dynamical modification may provide a way to induce changes in the brane structure, to impact strongly in the novel braneworld scenario to be described in the present study. For this, we organize our work as follows: in Sec. {\it 2} one introduces the general formalism that describes the braneworld model in the presence of several source scalar fields with generalized dynamics. Moreover, in
Sec. {\it3} one concentrates on the dynamics of cuscuton itself and exemplifies the formalism presented with two- and three-field models. We end the work in Sec. {\it4}, adding some conclusions and comments on future possibilities of new investigations.

{\it{2. Generalities.}} Let us start this investigation by considering a generalized $N$ field action in five-dimensional spacetime in the form
\be\label{action}
{\cal S}=\int d^4 x dy \sqrt{|g|} \left(-\frac{1}4R + {\cal L}_s\!\left(\phi_i,X_{ij}\right)\right),
\ee
where $i,j=1,2,\ldots,N$. Here we are using $4\pi G^{(5)}=1$
and $g=det(g_{ab})$ for $a, b=0, 1,\ldots,4$. We also define
the quantities $X_{ij}$ as
\be
X_{ij}=\frac12\nabla_a\phi_i\nabla^a\phi_j.
\ee

The equations of motion for the $N$ scalar fields can be written as
\be\label{Eqmov1}
\nabla_a\left({\cal L}_{X_{ij}}\nabla^a\phi_j\right)={\cal L}_{\phi_{i}},
\ee
where ${\cal L}_{X_{ij}}=\partial {\cal L}/\partial X_{ij}$ and ${\cal L}_{\phi_{j}}=\partial {\cal L}/\partial {\phi_{j}}$. Note that the Eq. \eqref{Eqmov1} describe $N$ coupled differential equations since the index $j$ is being summed for the $N$ fields. We can also obtain the Einstein equation in the usual form
\be\label{EinEq}
R_{ab}-\frac{1}2g_{ab}\,R =2T_{ab},
\ee
where the energy-momentum tensor is given by
\be
T_{ab}=\nabla_a\phi_i\nabla_b\phi_j{\cal L}_{X_{ij}}-g_{ab}\,{\cal L}\,.
\ee

We now concentrate on the study of static configurations. For this, let us assume that the fields only depend on the extra dimension, such that $\phi_i=\phi_i(y)$. Moreover, we consider the usual line element for the study of braneworld models, that is,
\be
 ds^2=e^{2A(y)}\eta_{\mu\nu}dx^\mu dx^\nu-dy^2\,,
\ee
where $\eta_{\mu\nu}\!=\!(1,\!-1,\!-1,\!-1)$ and $\exp(2A)$ is the warp factor, which we assume to depend only on the extra dimension, that is, $A=A(y)$. With this, the equations of motion for the scalar fields can be written as
\be\label{fielstati}
\begin{aligned}
\big({\cal L}_{X_{ij}}-{\cal L}_{X_{ik}X_{jl}}\phi'_{k}\phi'_{l}\big)\phi''_j+
{\cal L}_{\phi_{i}}\;\;\;\\
+\,\phi'_{j}\phi'_{k}{\cal L}_{X_{ij}\phi_{k}}+4{\cal L}_{X_{ij}}\phi'_{j}A'=0.
\end{aligned}
\ee
Here the prime denotes derivative with respect to the extra dimension. On the other hand, the two non-vanishing components of the Einstein equations \eqref{EinEq} become
\bes\label{Einstein}
\bal
A''&=-\frac23{\cal L}_{X_{ij}}\phi'_{i}\phi'_{j},\label{CompEinstati01}\\
A'^2&=\frac13\left({\cal L}+{\cal L}_{X_{ij}}\phi'_{i}\phi'_{j}\right)\label{CompEinstati02}.
\eal
\ees
In the case of static configurations, the energy density of the brane is obtained by $\rho(y)=T_{00}=-e^{2A}{\cal L}$. By using the Eqs.~\eqref{CompEinstati01} and \eqref{CompEinstati02} we get
\be
\rho(y)=-\frac32\left(e^{2A}A'\right)^{\prime}\,.
\ee

To deal with the system of second-order differential equations represented by Eqs. \eqref{fielstati} and \eqref{Einstein} we can introduce the first-order formalism obtained with an auxiliary function $W$ that depends on the $N$ fields $\phi_i$ in the form
\be\label{FOF1}
{\cal L}_{X_{ij}}\phi'_j=W_{\phi_i}\,,\qquad A'=-\frac23W.
\ee
By using the first-order formalism we can write the energy density of the brane as
\be
\rho(y)=\frac{d}{dy}\left(e^{2A}W\right),
\ee
Note that if $W$ is not a complicated asymptotically divergent function, the energy of the brane vanishes if $e^{2A}\to0$ when $y\to\pm\infty$.

{\it{2.1. Linear Stability.}}
The investigation of the linear stability is done in the usual way. In general, we start by introducing a perturbed metric tensor in terms of a new variable $z$ as
\be
ds^2=e^{2A}\Big(\!\big(\eta_{\mu\nu}+h_{\mu\nu}\left(x^\alpha,z\right)\!\big)dx^\mu dx^\nu-dz^2\Big),
\ee 
where $h_{\mu\nu}(x^\alpha,z)$ satisfies the transverse and traceless (TT) conditions $\partial^{\mu} h_{\mu\nu}\!=\!0$ and $h_{\mu}{}^{\mu}\!=\!0$. We also consider fluctuations in scalars fields: $\phi_i\!\to\!\phi_i(z)+\delta\phi_i\left(x^\alpha,z\right)$. In addition, we rewrite $h_{\mu\nu}(x^\alpha,z)=e^{ik\cdot x}e^{-3A(z)/2} H_{\mu\nu}(z)$. After some mathematical manipulations we get the linearization of the Einstein equation \eqref{EinEq} as
\be\label{Schrodinger}
\left(-\frac{d^2}{dz^2}+{\cal U}(z)\right)H_{\mu\nu}=k^2 H_{\mu\nu}\,,
\ee 
where
\be\label{potquant}
{\cal U}(z)=\frac32\frac{d^2\!A}{dz^2}+\frac94\left(\frac{dA}{dz}\right)^2\,.
\ee 
So, we see that the theory is stable by linear perturbations since Eq. \eqref{Schrodinger} can be written as $S^{\dagger}S H_{\mu\nu}=k^2H_{\mu\nu}$, where $S=-d/dz-(3/2)dA/dz$ and $k^2>0$. 

It is possible to show that we can get the zero mode, with $k=0$, in the variable $y$ as
\be
\xi_{\mu\nu}^{(0)}(y)=N_{\mu\nu}\,e^{2A(y)}\,,
\ee 
where $N_{\mu\nu}$ is a normalization. In the following examples we depict the potential \eqref{potquant} also in the variable $y$, which can be done by considering $dz=e^{-A}dy$.

{\it {3. Cuscuton-like dynamics.}}
Let us now consider a specific Lagrange density that engenders cuscuton-like dynamic in the form
\ben
\!\!\!{\cal L}_s=\!\sum_{i}\!\left(\frac12\nabla_a\phi_i\nabla^a\phi_i\!+\!\alpha_i \sqrt{|\nabla_a\phi_i\nabla^a\phi_i|}\right)\!-V,
\een
where $\alpha_i$ are non-negative parameters that control the cuscuton terms for each field $\phi_i$. In this case, the equations of motion for static configurations become
\ben
\phi_i''+4A'\left(\phi_i'-\alpha_i\right)=V_{\phi_i}.
\een
The components of the Einstein equations are
\bes
\bal
A''&=-\frac23\sum_{i}\left(\phi_i'^2-\alpha_i\phi_i'\right),\\
A'^2&=\frac16\sum_{i}\phi_i'^2-\frac13 V.
\eal
\ees
Moreover, the first-order formalism defined by the set of Eqs.~\eqref{FOF1} allows that we write the first order equations
\be\label{FOE}
\phi'_i=W_{\phi_i}+\alpha_i\,,\qquad A'=-\frac23W.
\ee

We can also represent the scalar potential $V$ of the brane as
\be\label{potcalar}
V=\frac12\sum_{i}\left(W_{\phi_i}+\alpha_i\right)^2-\frac43W^2.
\ee
Here we notice that the term $W^2$ in Eq. \eqref{potcalar} is responsible for coupling the fields, even when $W$ is written as a sum of terms, each one depending on a single field, as we consider below. Let us now proceed with the study and specify a function $W(\phi_i)$ in the form
\be\label{ModP4}
W(\phi_i)=\!\sum_{i}\beta_i\!\left(\!\phi_i-\frac13\phi_i^3\!\right),
\ee 
where $\beta_i$ are real non-negative parameters that adjust the profile and coupling of the fields. By using the Eq.~\eqref{ModP4} in the first of Eqs. \eqref{FOE}, we get the following set of first-order equations that describe the dynamics of the fields
\be
\phi_i'=\beta_i\left(1-\phi_i^2\right)+\alpha_i\,.
\ee
We can solve this differential equation analytically to obtain solutions of the fields that can be represented as
\be
\phi_i(y)=\sqrt{\frac{\alpha_i\!+\!\beta_i}{\beta_i}}\,\tanh\Big(\sqrt{\beta_i(\alpha_i\!+\!\beta_i)}\,(y+c_i)\Big),
\ee
where $c_i$ are real constants which describe the centers of the kink-like solutions. We can now use the second of the Eqs.~\eqref{FOE} to obtain the warp function as 
\be
A(y)=\sum_{i}A_i(y)+A_0,
\ee
where $A_0$ is the constant of integration introduced to make $A(0)=0$ and $A_i(y)$ is defined as
\be
\begin{aligned}
A_i(y)=&\,\frac{4\beta_i\!-\!2\alpha_i}{9\beta_i}\ln\!\left(\sech\big(\!\sqrt{\!\beta_i(\alpha_i\!+\!\beta_i)}\,(y\!+\!c_i)\big)\right)\\
&\!+\!\frac{\beta_i\!+\!\alpha_i}{9\beta_i}\sech^2\big(\!\sqrt{\!\beta_i(\alpha_i\!+\!\beta_i)}\,(y\!+\!c_i)\big).
\end{aligned}
\ee
It is possible to show that asymptotically the warp function behaves as
\ben
A(y\gg0)\approx-\frac29\sum_{i}\sqrt{\frac{\alpha_i\!+\!\beta_i}{\beta_i}}\,\,(2\beta_i\!-\!\alpha_i)\,\big|y\big|.
\een
Also, the five-dimensional cosmological constant is
\ben
\Lambda_5\equiv V(\phi_i)=-\frac4{27}\sum_{i}\frac{\alpha_i\!+\!\beta_i}{\beta_i}\big(2\beta_i\!-\!\alpha_i\big)^2.
\een
These results show that under adequate choices of $\alpha_i$ and $\beta_i$, the warp factor may have the desired profile.

\begin{figure}[t]
    \begin{center}
        \includegraphics[scale=0.6]{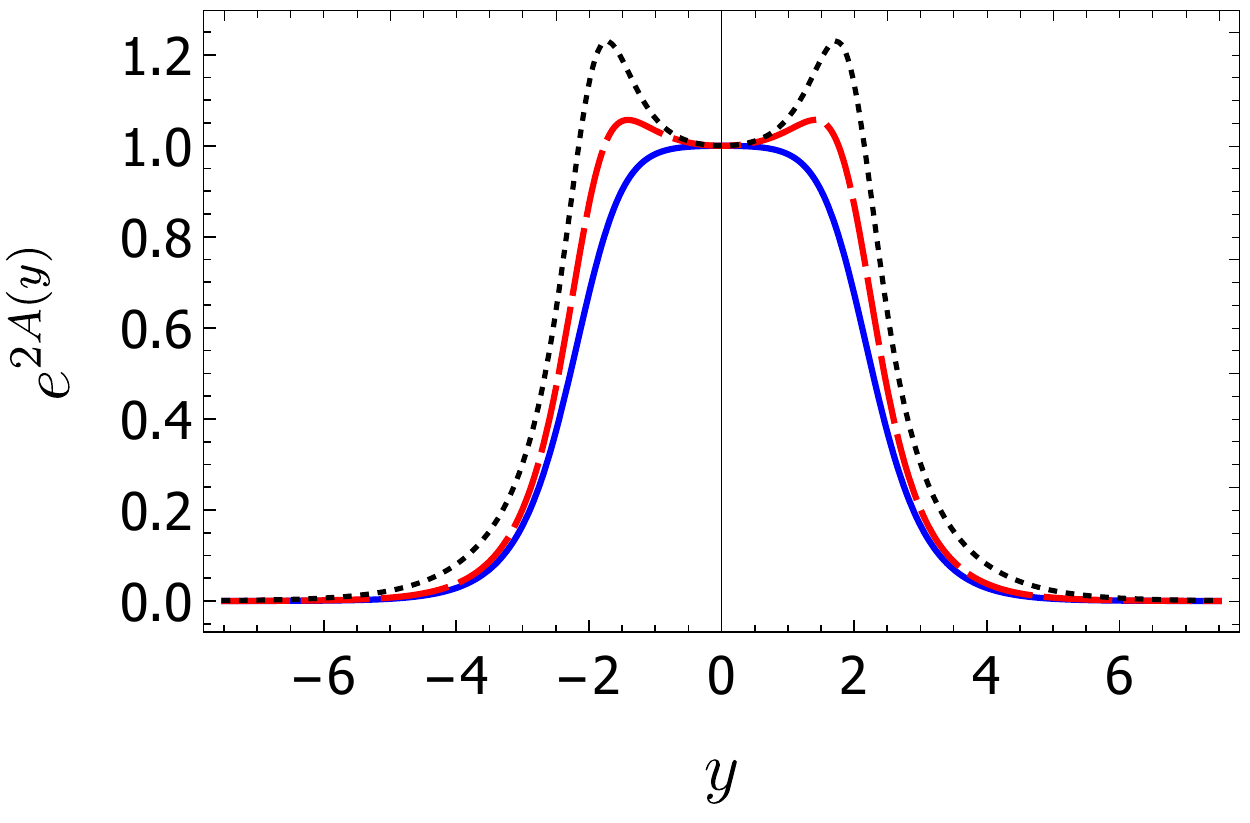}
        \includegraphics[scale=0.6]{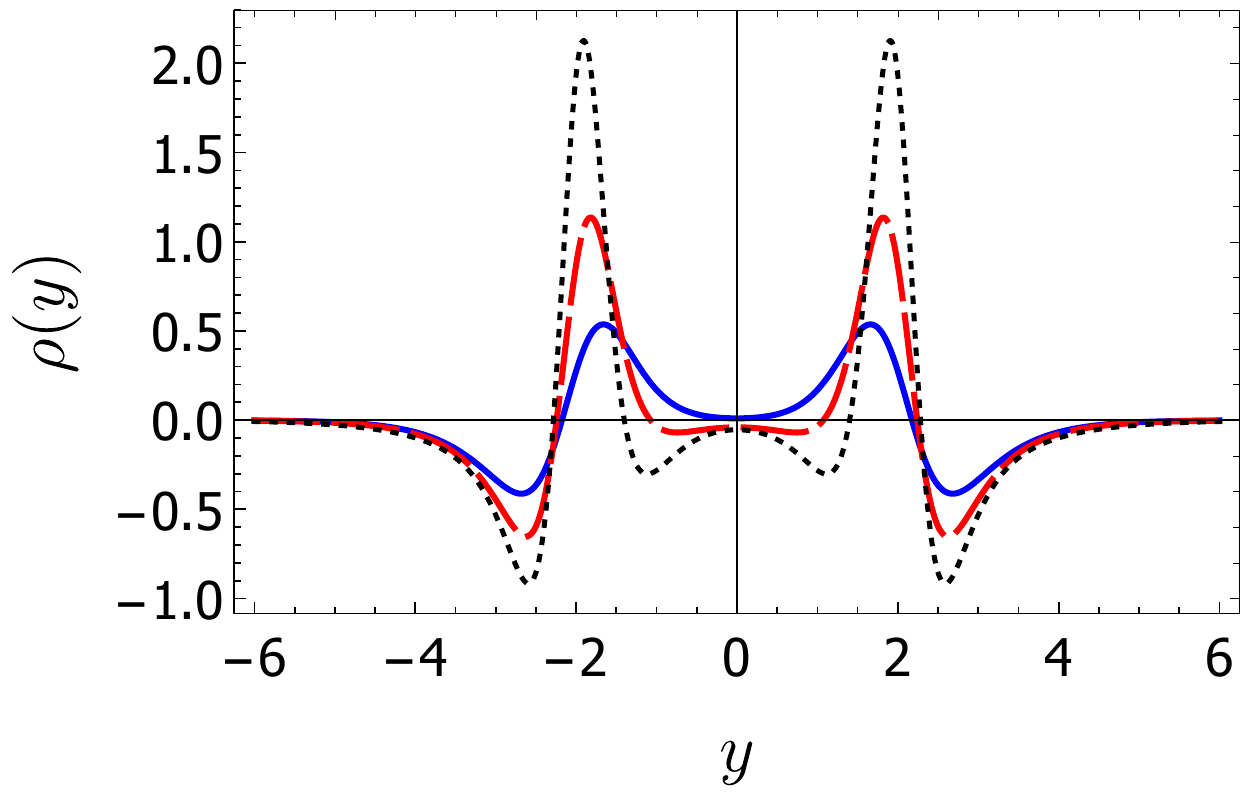}
    \end{center}
    \vspace{-0.5cm}
    \caption{\small{Warp factor (top panel) and energy density (bottom panel) depicted for $c=2$ and $\alpha=0$ (solid-line), $\alpha=0.5$ (dashed-line) and $\alpha=1$ (dotted-line).}\label{fig1}}
\end{figure}

{\it{3.1. Two-field model.}}
We now investigate more specific situations, first considering a two-field model where the cuscuton term acts equally for both. In this case we assume that $\alpha_i=\alpha$. We have noticed that considering different $\alpha_i$ only generates an asymmetry in the quantities studied. The $\beta_i$ parameters acts in a similar way, so we assume for simplicity that $\beta_i=1$. Furthermore, we follow Ref. \cite{Bazeia:2022BL} and consider that the two solutions are displaced symmetrically around the origin. We can then write the solutions of the two fields as
\bes\label{soluMA}
\bal
\phi_1(y)&=\sqrt{\alpha\!+\!1}\,\tanh\big(\sqrt{\alpha\!+\!1}\,(y+c)\big),\\
\phi_2(y)&=\sqrt{\alpha\!+\!1}\,\tanh\big(\sqrt{\alpha\!+\!1}\,(y-c)\big).
\eal
\ees
In this case, we get the warp function as
\be\label{warpFacMA}
\begin{aligned}
A(y)=&\,\frac{4\!-\!2\alpha}{9}\ln\!\Big(\sech\big(\!\sqrt{\alpha\!+\!1}\,(y\!+\!c)\big)\Big)\\
&\!+\!\frac{4\!-\!2\alpha}{9}\ln\!\Big(\sech\big(\!\sqrt{\alpha\!+\!1}\,(y\!-\!c)\big)\Big)\\
&\!+\!\frac{1\!+\!\alpha}{9}\sech^2\big(\!\sqrt{\alpha\!+\!1}\,(y\!+\!c)\big)\\
&\!+\!\frac{1\!+\!\alpha}{9}\sech^2\big(\!\sqrt{\alpha\!+\!1}\,(y\!-\!c)\big)+A_0,
\end{aligned}
\ee
where
\be
\begin{aligned}
A_0=&\,-\frac{8-4\alpha}{9}\ln\left(\sech\big(\sqrt{\alpha+1}\,c\big)\right)\\
&\,-\frac{2+2\alpha}{9}\,\sech^2\big(\sqrt{\alpha+1}\,c\big).
\end{aligned}
\ee
In this case we have to use $\alpha \in[0,2)$.  In the top panel of the Fig. \ref{fig1} we shows the warp factor $\exp(2A)$ using the Eq. \eqref{warpFacMA} with $c=2$ and $\alpha=0$ (solid-line), $\alpha=0.5$ (dashed-line) and $\alpha=1$ (dotted-line). As we can see, the cuscuton works to split the warp factor, introducing an interesting modification in the internal structure of the gravitational sector of the brane. This result remind us of the one obtained in \cite{Bazeia:2021jok}  where the authors also studied cuscuton dynamics in a similar scenario. In the present study we have found that no split in warp factor is observed for $\alpha=0$, i.e., without cuscuton dynamics, which agrees with the previous work \cite{Bazeia:2022BL}. As can be seen, the cuscuton dynamics seems to exert an important influence on the internal structure of the warp factor. We investigated the Kretschmann scalar, obtained by $K(y)=40A'^4+16A''^2+32A'^2A''$ and verified that it behaves properly for the parameters used in this study.

The energy density for this model can be described analytically by using the solutions \eqref{soluMA} and Eq.~\eqref{warpFacMA}. Instead of presenting the complete equation from the energy density, we prefer to just represent its behavior as shown in the bottom panel of the Fig. \ref{fig1}. Note that the energy density is also influenced by the cuscuton term, for example, having the point at $y=0$ changing between a minimum and a maximum.

\begin{figure}[t]
    \begin{center}
        \includegraphics[scale=0.6]{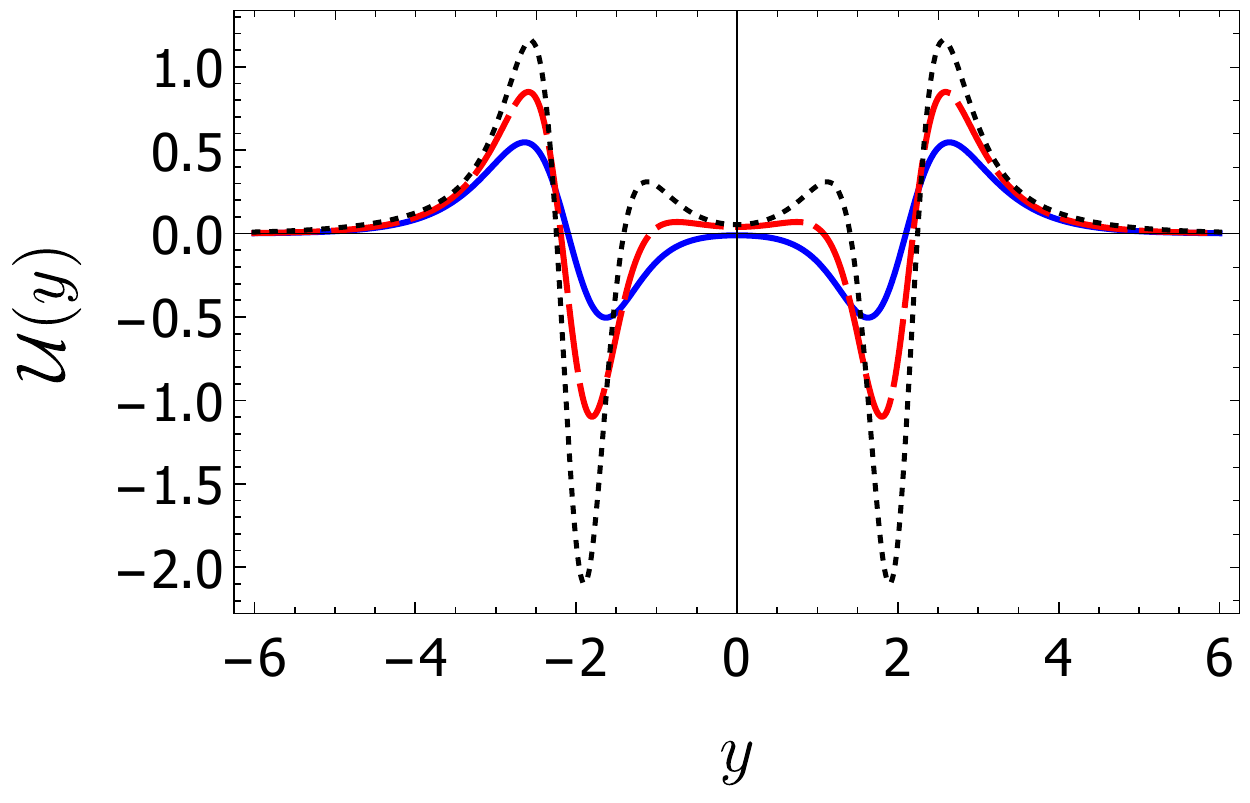}
        \includegraphics[scale=0.6]{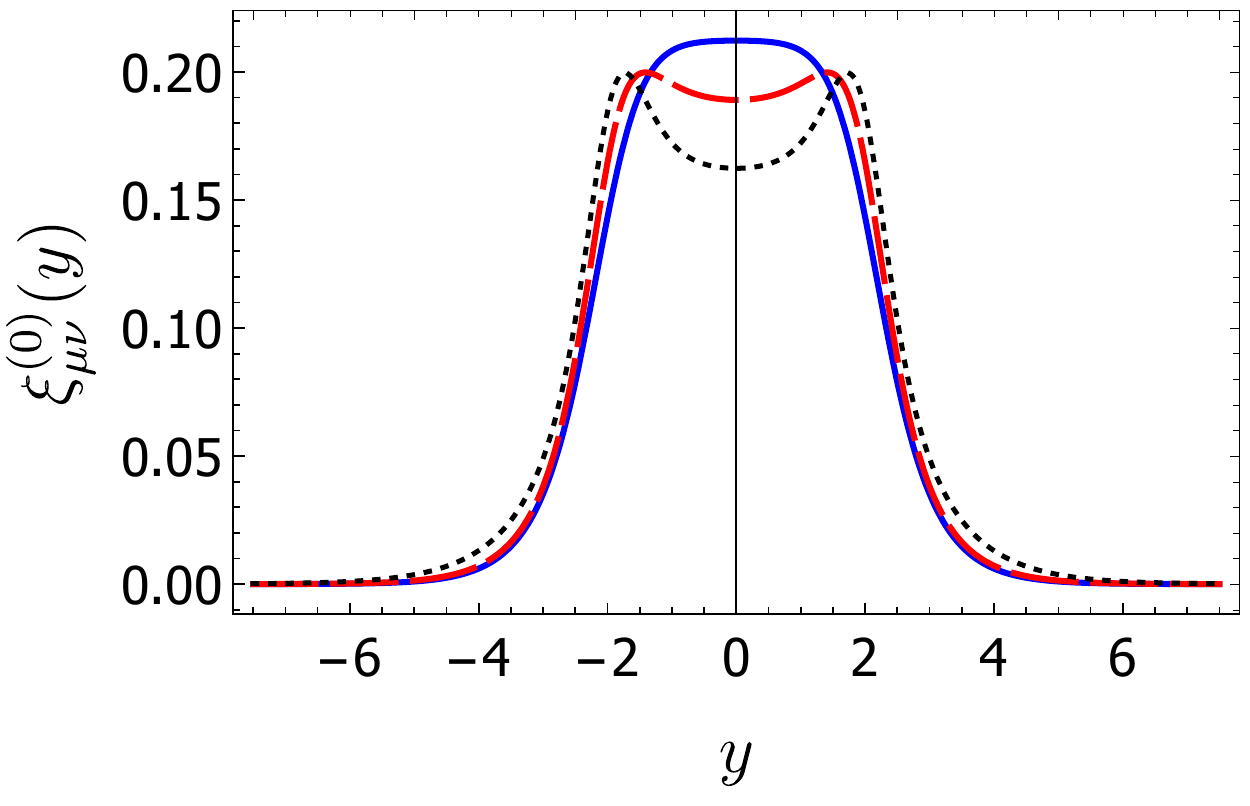}
    \end{center}
    \vspace{-0.5cm}
    \caption{\small{Stability potential (top panel) and zero mode (bottom panel) depicted for $c=2$ and $\alpha=0$ (solid-line), $\alpha=0.5$ (dashed-line) and $\alpha=1$ (dotted-line).}\label{fig2}}
\end{figure}

Finally, we also check the stability potential and the zero mode as can be seen in the top and bottom panels of the Fig. \ref{fig2}, respectively. As we can see, the cuscuton term also causes changes here. The stability potential undergoes an intense change in behavior, being able to present two displaced minima with a central maximum or a central minimum surrounded by two other minima. In addition, the zero mode acquires an interesting internal structure, which was not observed in previous works. 

Several other possibilities can be explored for the change of the parameters $\alpha_i$ and $\beta_i$. For example, we can obtain asymmetric brane solutions by either introducing different $\alpha_i$ to control the cuscuton term of each field, or providing different amplitudes and thickness for the solutions by changing the parameters $\beta_i$. 

{\it{3.2. Three-field model.}}
Let us now consider a three-field model. Again, for simplicity, we assume that $\beta_i=1$ and that all the $\alpha_i$ are identical to $\alpha$. Moreover, we keep one of the three fields centered at $y=0$, and the other two symmetrically located around the origin. Furthermore, we consider two different situations; in the first, only the central field will have cuscuton dynamics; in the second, only the central field will not have cuscuton dynamics.

{\it{3.2.1. First case.}}
Considering that only the central field has cuscuton dynamics, we can write the solutions as
\bes
\bal
\phi_1(y)&=\sqrt{\alpha\!+\!1}\,\tanh\big(\sqrt{\alpha\!+\!1}\,y\big),\\
\phi_2(y)&=\tanh\big(y+c\big),\\
\phi_3(y)&=\tanh\big(y-c\big).
\eal
\ees

\begin{figure}[t]
    \begin{center}
        \includegraphics[scale=0.6]{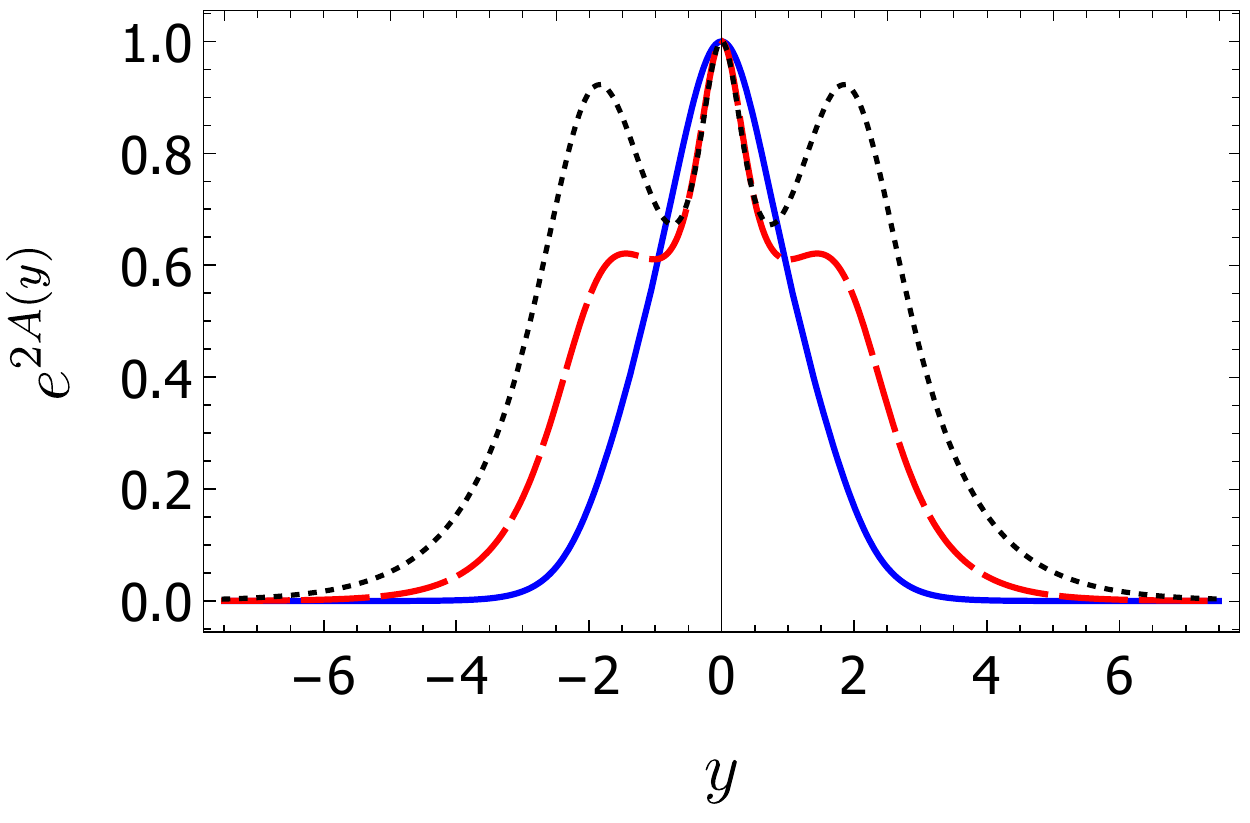}
        \includegraphics[scale=0.6]{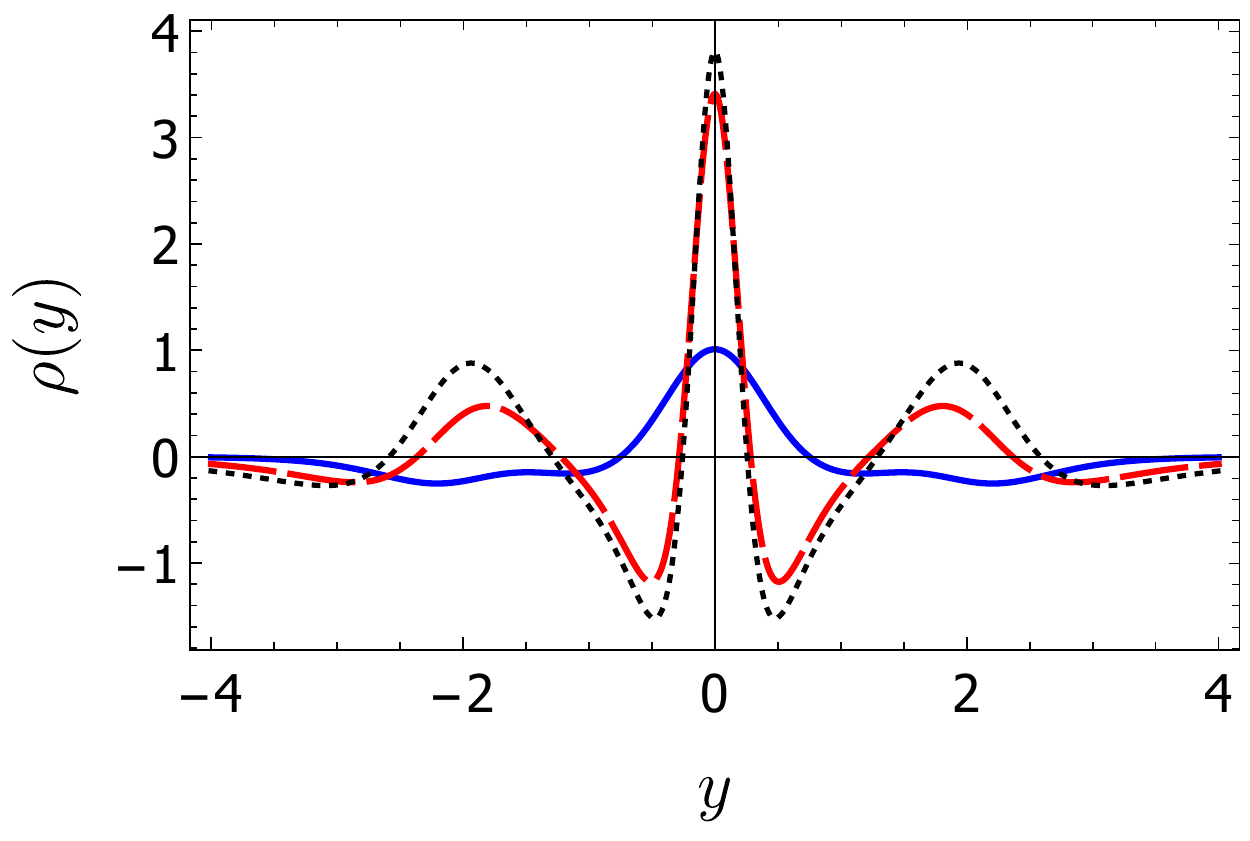}
    \end{center}
    \vspace{-0.5cm}
    \caption{\small{Warp factor (top panel) and energy density (bottom panel) depicted for $c=2$ and $\alpha=0$ (solid-line), $\alpha=2.4$ (dashed-line) and $\alpha=2.8$ (dotted-line).}\label{fig3}}
\end{figure}

In this case the warp function assume the form
\be\label{warpFacMB1}
\begin{aligned}
A(y)=&\,\frac{4\!-\!2\alpha}{9}\ln\!\Big(\sech\big(\!\sqrt{\alpha\!+\!1}\,y\big)\Big)\\
&\!+\!\frac{4}{9}\ln\!\Big(\sech(y\!-\!c)\,\sech(y\!+\!c)\Big)\\
&\!+\!\frac{1}{9}\sech^2\big(y\!+\!c\big)\!+\!\frac{1}{9}\sech^2\big(y\!-\!c\big)\\
&\!+\!\frac{1\!+\!\alpha}{9}\sech^2\big(\!\sqrt{\alpha\!+\!1}\,y\big)+A_0,
\end{aligned}
\ee
where
\ben
A_0=\,-\frac{8}{9}\ln\!\big(\sech(c)\big)\!-\!\frac{2}{9}\sech^2(c)\!-\!\frac{1\!+\!\alpha}{9}\,.
\een
Here we must consider $\alpha\in[0,3.82)$. Fig. \ref{fig3} shows the warp factor $\exp(2A)$ obtained by \eqref{warpFacMB1}. In this figure we use $c=2$ and $\alpha=0$ (solid-line), $\alpha=2.4$ (dashed-line) and $\alpha=2.8$ (dotted-line). Here we obtain a new configuration for the warp factor, where the central maximum is surrounded by two other peaks, indicating a new representation for the internal structure of the brane that, as far as we know, had not been suggested before. We also checked the Kretschmann scalar and obtained that it behaves properly, despite the intense change in the warp factor. This indicates that the modifications introduced in this study are robust, as they do not appear to introduce any undesirable change to the brane scenario.

The energy density for this new situation is represented in the bottom panel of Fig. \ref{fig3}. Here the changes are not dramatic, since the representation is quite similar to the usual case without cuscuton. See that the displacement attributed to the location of the centers of two of the three fields of the model induces a greater spread of the energy density around the origin, specifically around the points $y=\pm2$.

\begin{figure}[t]
    \begin{center}
        \includegraphics[scale=0.6]{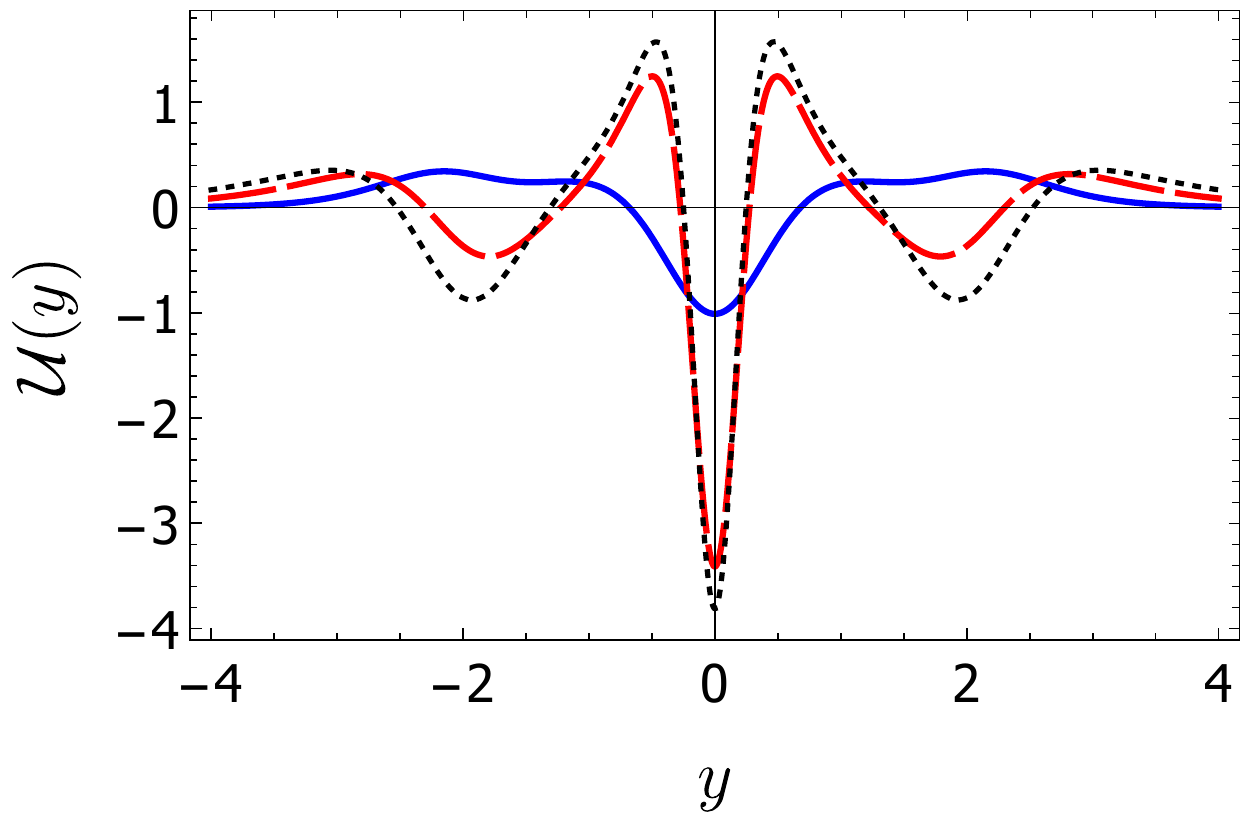}
        \includegraphics[scale=0.6]{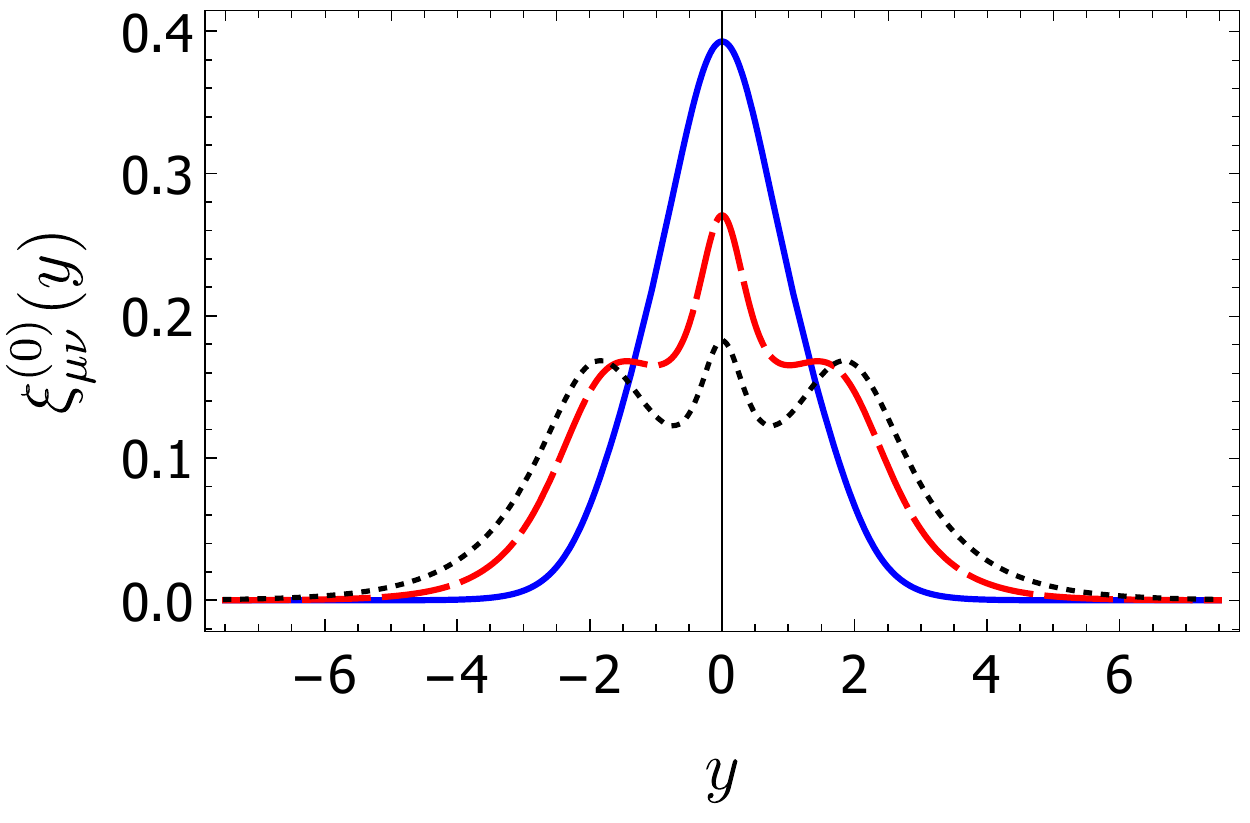}
    \end{center}
    \vspace{-0.5cm}
    \caption{\small{Stability potential (top panel) and zero mode (bottom panel) depicted for $c=2$ and $\alpha=0$ (solid-line), $\alpha=2.4$ (dashed-line) and $\alpha=2.8$ (dotted-line).}\label{fig4}}
\end{figure}

We found that the stability potential is also affected by this change as can be seen in the Fig. \ref{fig4}. In this case, the cuscuton term acts to support new minima in the potential, which may induce the appearance of new bound states. The zero mode feels the change in the stability potential and starts to present a completely different internal structure from what is obtained in more usual situations as can be seen in the bottom panel of the Fig.~\ref{fig4}. In fact, we have presented a new behavior, which differs substantially from other results presented in previous braneworld scenarios.

\begin{figure}[tp]
    \begin{center}
        \includegraphics[scale=0.6]{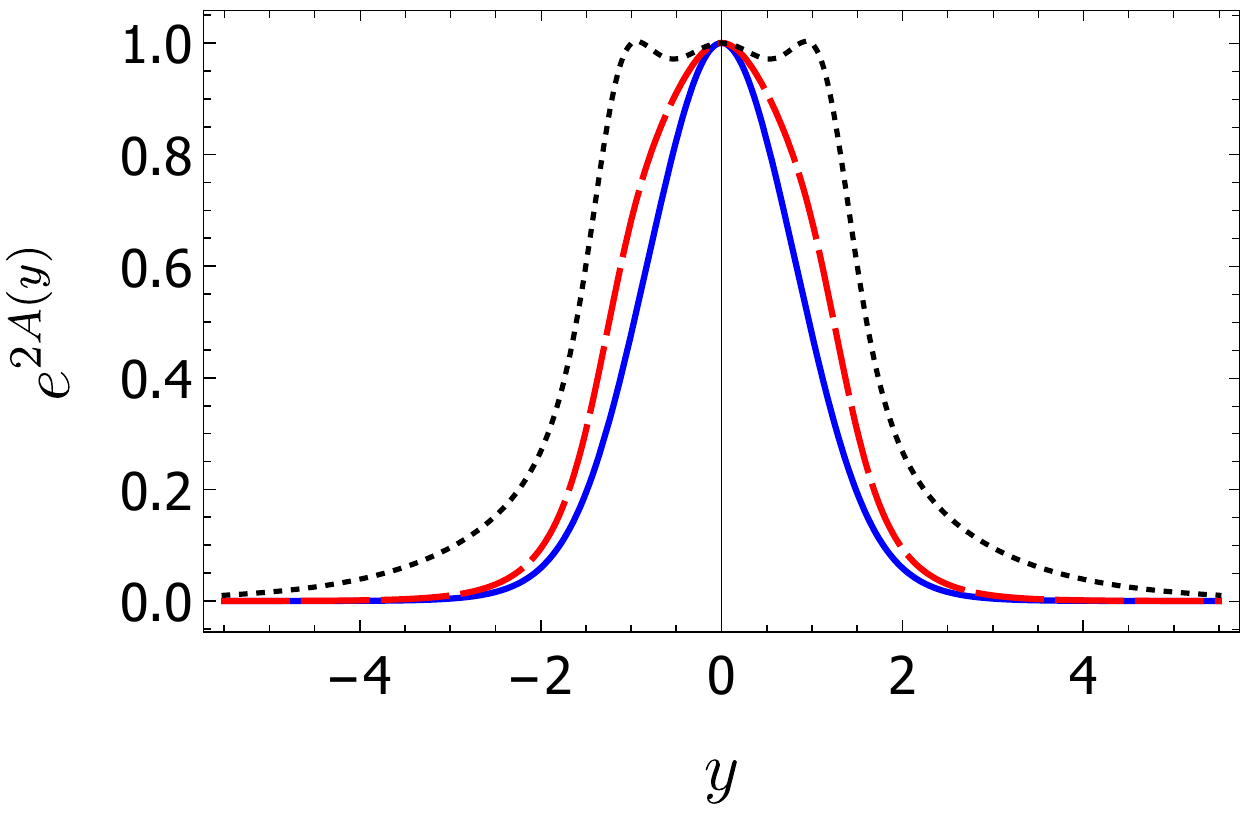}
        \includegraphics[scale=0.6]{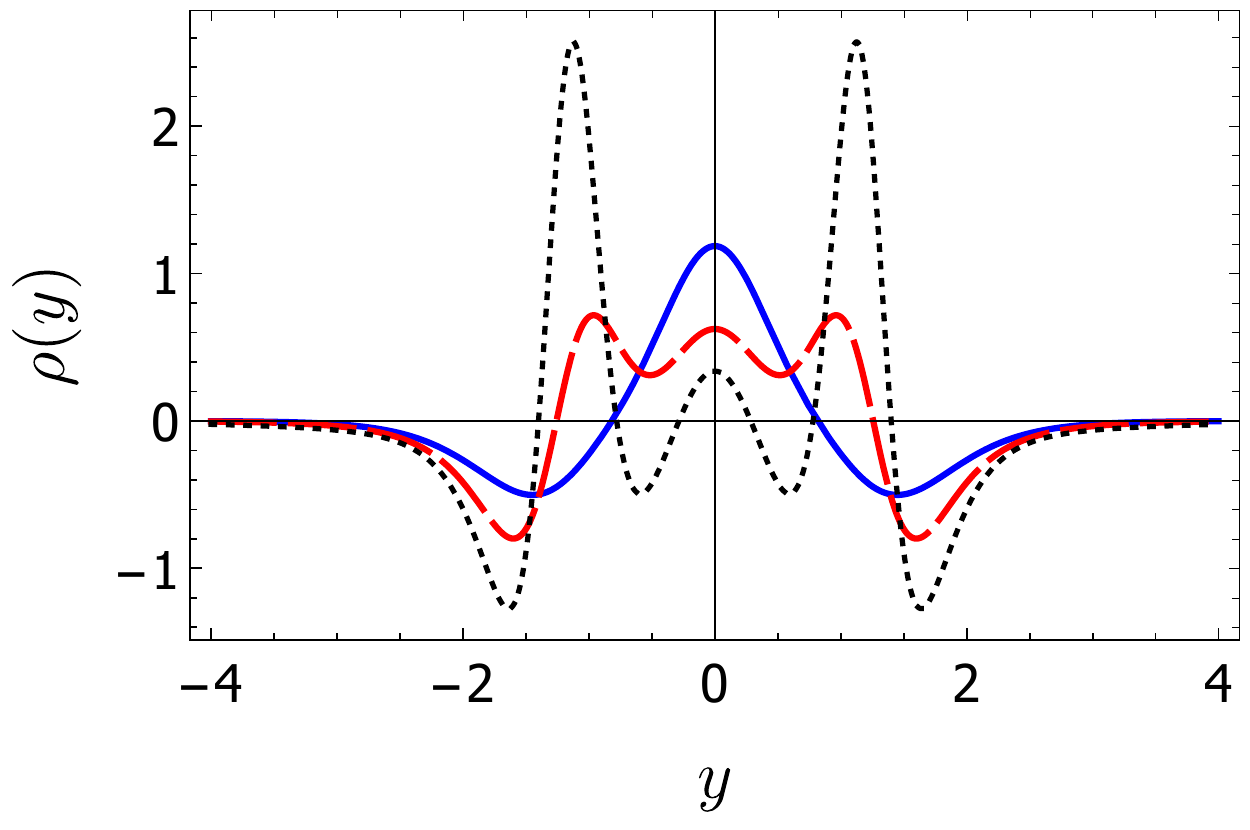}
    \end{center}
    \vspace{-0.5cm}
    \caption{\small{Warp factor (top panel) and energy density (bottom panel) depicted for $c=1.2$ and $\alpha=0$ (solid-line), $\alpha=1$ (dashed-line) and $\alpha=2$ (dotted-line).}\label{fig5}}
\end{figure}

{\it{3.2.2. Second case.}}
Let us now consider a different situation and suppose the central field has standard dynamics, but the others two have the cuscuton contribution. In this case, the solutions are
\bes
\bal
\phi_1(y)&=\tanh(y),\\
\phi_2(y)&=\sqrt{\alpha\!+\!1}\,\tanh\big(\sqrt{\alpha\!+\!1}\,(y+c)\big),\\
\phi_3(y)&=\sqrt{\alpha\!+\!1}\,\tanh\big(\sqrt{\alpha\!+\!1}\,(y-c)\big).
\eal
\ees
Then, the warp function is
\be\label{warpFacMB2}
\begin{aligned}
A(y)=&\,\,\frac{2}{9}(2-\alpha)\ln\!\Big(\sech\big(\!\sqrt{\alpha\!+\!1}\,(y+c)\big)\Big)\\
&\!+\!\frac{2}{9}(2-\alpha)\ln\!\Big(\sech\big(\!\sqrt{\alpha\!+\!1}\,(y-c)\big)\Big)\\
&\!+\!\frac{1}{9}(1+\alpha)\,\sech^2\big(\!\sqrt{\alpha\!+\!1}\,(y+c)\big)\\
&\!+\!\frac{1}{9}(1+\alpha)\,\sech^2\big(\!\sqrt{\alpha\!+\!1}\,(y-c)\big)\\
&\!+\!\frac{4}{9}\ln\!\big(\sech(y)\big)+\frac{1}{9}\sech^2(y)+A_0,
\end{aligned}
\ee
and now
\be
\begin{aligned}
A_0=&\,-\frac49(2-\alpha)\ln\!\Big(\sech\big(\sqrt{1\!+\!\alpha}\,c\big)\Big)\\
&\,-\frac29(1+\alpha)\,\sech^2\big(\sqrt{1\!+\!\alpha}\,c\big)-\frac{1}{9}\,.
\end{aligned}
\ee
In this case we must consider $\alpha\in [0,2.53)$. Fig. \ref{fig5} shows the warp factor obtained by Eq. \eqref{warpFacMB2} with $c=1.2$ and $\alpha=0, 1$ and $2$. Once again, we see that the warp factor presents different behaviors depending on the choice of the $\alpha$ and $c$ parameters, and may have one or three maxima. The bottom panel of the Fig. \ref{fig5} represents the energy density for the same values as the parameters used to represent the warp factor.

\begin{figure}[tp]
    \begin{center}
        \includegraphics[scale=0.6]{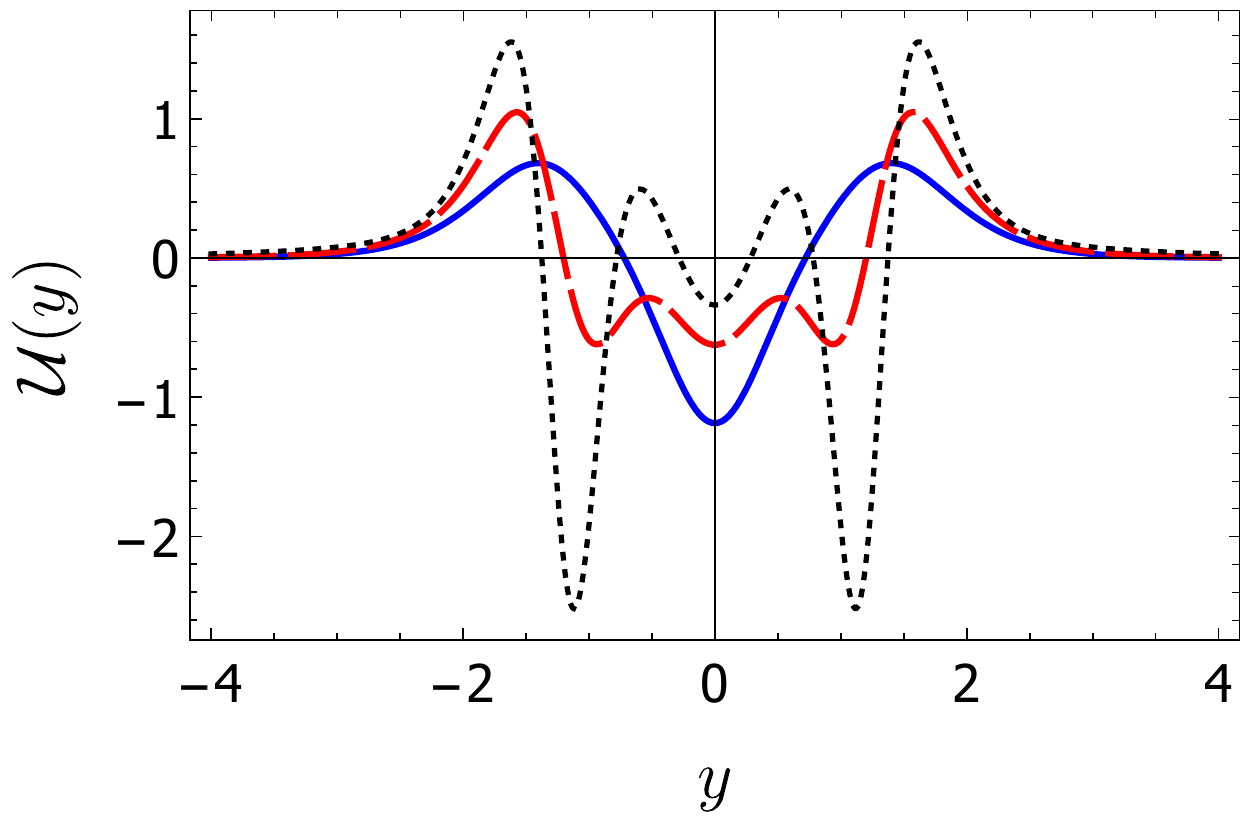}
        \includegraphics[scale=0.6]{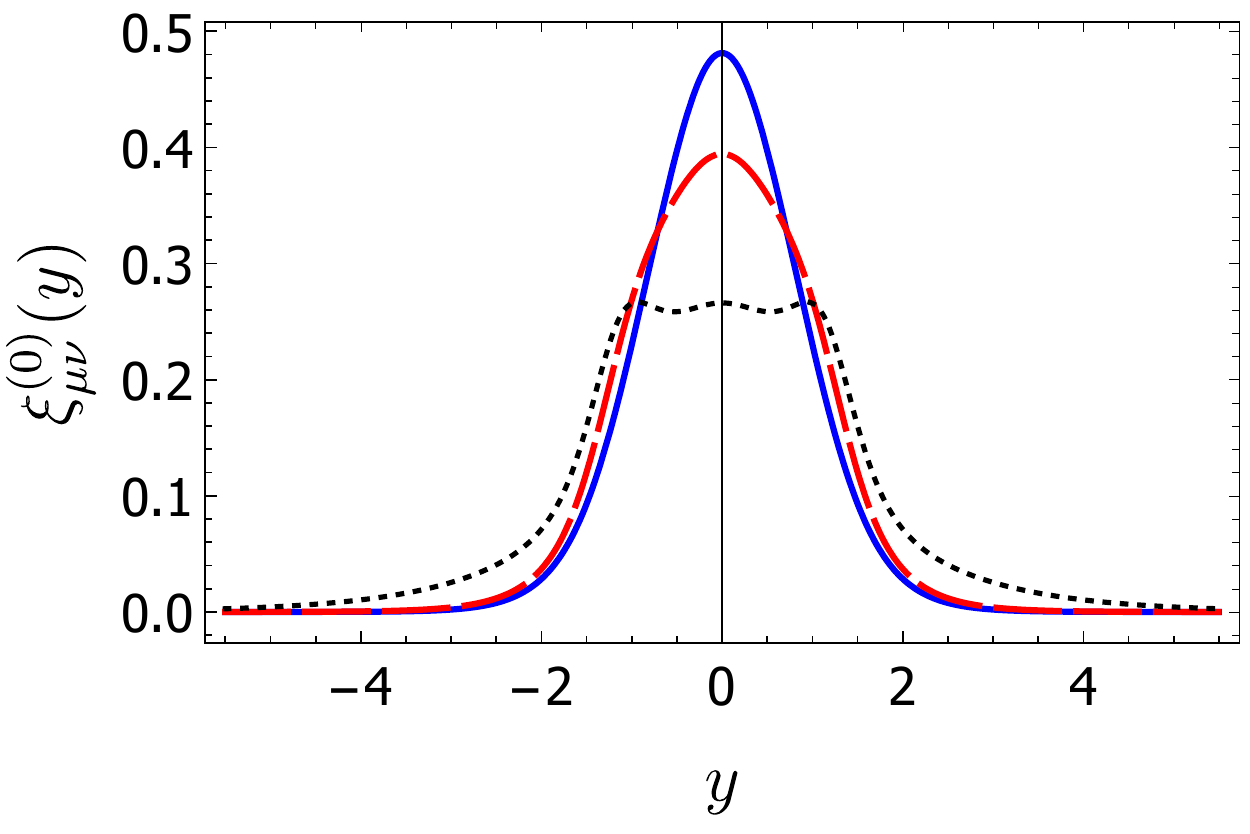}
    \end{center}
    \vspace{-0.5cm}
    \caption{\small{Stability potential (top panel) and zero mode (bottom panel) depicted for $c=1.2$ and $\alpha=0$ (solid-line), $\alpha=1$ (dashed-line) and $\alpha=2$ (dotted-line).}\label{fig6}}
\end{figure}
The stability potential and zero mode are represented in the top and bottom panels of the Fig. \ref{fig6}, respectively. Once again, we get a unique internal structure for the mode zero. Several other possibilities can also be considered. For example, we could assume different $\alpha_i$ to control the cuscuton term of each field, which should generate an asymmetry in the system. Furthermore, we could consider $\alpha$ or $\beta$ different only for the central field, in which case it is verified that the central region can be adjusted in order to enlarge or reduce the minima or maxima around the center at $y=0$.

{\it{4. Ending comments.}}
In this work we investigated braneworld models in the multi-field scenario, molded by multiple scalar fields with generalized dynamics. We followed the proposal suggested in \cite{Bazeia:2006ef,Ahmed:2012nh,deSouzaDutra:2014ddw} and considered a polynomial model where the potential of the brane couples the several scalar fields of the theory. In this sense, we analyzed situations with two and three fields, with at least one of them engendering cuscuton-like dynamics. Moreover, we used the mechanism presented in \cite{Bazeia:2022BL} to control the internal structure of the solutions.  In each case, we verified how the cuscuton term contributed to modify the warp factor, energy density, stability potential and the zero mode associated to the braneworld configurations.

We started the investigation considering a two-field model where both fields had cuscuton dynamics. We followed \cite{Bazeia:2022BL} and varied the cuscuton parameter in order to understand its influence on the model. We found that the cuscuton term contributed to altering the internal structure of the brane, causing a split in the warp factor. Because of this, we also noticed modifications in the energy density and stability potential. In particular, the stability potential may engender a central maximum or minimum surrounded by two wells that increase in depth as one enlarges the parameter that controls the cuscuton term. Furthermore, the zero mode loads the division of the warp factor, indicating to be more concentrated in the stability potential wells that are influenced by the cuscuton parameter.

We also analyzed a model with three fields in two different situations, keeping the center of the third field centered at the origin. First, we considered that only the central field had cuscuton dynamics. In this case, we identified an interesting change in the internal structure of the warp factor, allowing the emergence of one, two or three peaks depending on the values of the parameters used. As far as we know, this form of internal structure in the warp factor has never been obtained before, which is an interesting consequence of introducing cuscuton dynamics into the model. As expected, significant changes also arose in energy density, stability potential and zero mode in this case. We also analyzed the situation where only the solutions of the fields that are displaced from the center had cuscuton dynamics. In this case, we found results similar to the previous ones, with the solutions having an equally rich and distinct internal structure.

In view of what was presented in this study, we found that the cuscuton dynamics contributes in a non-negligible way to the emergence of internal structures in braneworld models with multiple scalar fields. We have found a new and interesting way of controlling the internal structure of brane, which can, for example, change how gravity is trapped in the brane. The modification of the internal structure can induce interesting changes in the study of other issues, for example, in the case of mimetic gravity \cite{Zhong:2017uhn,Zhong:2018fdq} and when we couple fermions to the brane \cite{Almeida:2009jc,Castro:2010uj,Moreira:2021vcf}. Furthermore, the investigation presented here can be used in other contexts of current interest such as the domain-wall/brane-cosmology correspondence \cite{DWAA,DWBB} and also in gravity in two dimensions, related to the Jackiw-Teitelboim and Karch-Randall gravity/braneworld models \cite{Zhong:2021voa,Geng:2022tfc,Geng:2022slq}, in the case of the scalar-tensor representation of $f(R,T)$ braneworld models \cite{Rosa:2021tei,Rosa:2022fhl}, and in the study of cuscuton inflation, to verify the impact of cuscuton models on standard scalar field inflation \cite{Bartolo:2021wpt}. Some of the above issues are presently under investigation, and we hope to report on them in the near future.

{\it Declaration of competing interest:} The authors declare that they have no known competing financial interests or personal relationships that could have appeared to influence this work.

{\it Data availability:} This manuscript describes a theoretical work with no associated data to be deposited.

{\it Acknowledgments:} The work is partially financed by Paraiba State Research Foundation, FAPESQ-PB, grant No. 0015/2019. DB also thanks CNPq (Brazil), grant No. 303469/2019-6, for partial financial support.



\begin{thebibliography}{99}

\bibitem{Randall:1999vf}
L.~Randall and R.~Sundrum,
Phys. Rev. Lett. {83}, 4690 (1999).

\bibitem{Goldberger:1999uk}
W.~D.~Goldberger and M.~B.~Wise,
Phys. Rev. Lett. 83, 4922 (1999).

\bibitem{Skenderis:1999mm}
K.~Skenderis and P.~K.~Townsend,
Phys. Lett. B 468, 46 (1999).

\bibitem{DeWolfe:1999cp}
O.~DeWolfe, D.~Z.~Freedman, S.~S.~Gubser and A.~Karch,
Phys. Rev. D 62, 046008 (2000).

\bibitem{Csaki:2000fc}
C.~Csaki, J.~Erlich, T.~J.~Hollowood and Y.~Shirman,
Nucl. Phys. B 581, 309 (2000).

\bibitem{Brito:2001hd}
F.~Brito, M.~Cvetic and S.~Yoon,
Phys. Rev. D 64, 064021 (2001).

\bibitem{Campos:2001pr}
A.~Campos,
Phys. Rev. Lett. 88, 141602 (2002).

\bibitem{Kobayashi:2001jd}
S.~Kobayashi, K.~Koyama and J.~Soda, 
Phys. Rev. D \textbf{65}, 064014 (2002).

\bibitem{Bazeia:2004dh}
D.~Bazeia and A.~R.~Gomes,
JHEP 05, 012 (2004).

\bibitem{Koley:2005nv}
R.~Koley and S.~Kar,
Phys. Lett. B 623, 244 (2005).
[erratum: Phys. Lett. B 631, 199 (2005)].

\bibitem{Pal:2007gp}
S.~Pal and S.~Kar,
Gen. Rel. Grav. 41, 1165 (2009).

\bibitem{Bazeia:2008zx}
D.~Bazeia, A.~R.~Gomes, L.~Losano and R.~Menezes,
Phys. Lett. B 671, 402 (2009).

\bibitem{Liu:2009ega}
Y.~X.~Liu, Y.~Zhong and K.~Yang,
EPL 90, 51001 (2010).

\bibitem{Bazeia:2013euc}
D.~Bazeia, A.~S.~Lobao, L.~Losano and R.~Menezes,
Phys. Rev. D 88, 045001 (2013).

\bibitem{Zhong:2012nt}
Y.~Zhong and Y.~X.~Liu,
Phys. Rev. D 88, 024017 (2013).

\bibitem{Afonso:2007gc}
V.~I.~Afonso, D.~Bazeia, R.~Menezes and A.~Y.~Petrov,
Phys. Lett. B 658, 71 (2007).

\bibitem{Dzhunushaliev:2009dt}
V.~Dzhunushaliev, V.~Folomeev, B.~Kleihaus and J.~Kunz,
JHEP 04, 130 (2010).

\bibitem{Zhong:2010ae}
Y.~Zhong, Y.~X.~Liu and K.~Yang,
Phys. Lett. B 699, 398 (2011).

\bibitem{Liu:2011wi}
Y.~X.~Liu, Y.~Zhong, Z.~H.~Zhao and H.~T.~Li,
JHEP 06, 135 (2011).

\bibitem{Yang:2012hu}
J.~Yang, Y.~L.~Li, Y.~Zhong and Y.~Li,
Phys. Rev. D 85, 084033 (2012).

\bibitem{Menezes:2014bta}
R.~Menezes,
Phys. Rev. D 89, 125007 (2014).

\bibitem{Guo:2015qbt}
W.~D.~Guo, Q.~M.~Fu, Y.~P.~Zhang and Y.~X.~Liu,
Phys. Rev. D 93, 044002 (2016).

\bibitem{Charmousis:2002rc}
C.~Charmousis and J.~F.~Dufaux,
Class. Quant. Grav. 19, 4671 (2002).

\bibitem{Kofinas:2003rz}
G.~Kofinas, R.~Maartens and E.~Papantonopoulos,
JHEP 10, 066 (2003).

\bibitem{Cho:2001nf}
Y.~M.~Cho and I.~P.~Neupane,
Int. J. Mod. Phys. A 18, 2703 (2003).

\bibitem{Maeda:2003vq}
Kei-ichi Maeda and T.~Torii,
Phys. Rev. D 69, 024002 (2004).

\bibitem{Afshordi:2006ad}
N.~Afshordi, D.~J.~H.~Chung and G.~Geshnizjani,
Phys. Rev. D 75, 083513 (2007).

\bibitem{Afshordi:2007yx}
N.~Afshordi, D.~J.~H.~Chung, M.~Doran and G.~Geshnizjani,
Phys. Rev. D 75, 123509 (2007).

\bibitem{Bazeia:2012br}
D.~Bazeia, F.~A.~Brito and F.~G.~Costa,
Phys. Rev. D 87, 065007 (2013).

\bibitem{Iyonaga:2018vnu}
A.~Iyonaga, K.~Takahashi and T.~Kobayashi,
JCAP 12, 002 (2018).

\bibitem{Ito:2019fie}
A.~Ito, A.~Iyonaga, S.~Kim and J.~Soda,
Phys. Rev. D 99, 083502 (2019).

\bibitem{Andrade:2018afh}
I.~Andrade, M.~A.~Marques and R.~Menezes,
Nucl. Phys. B 942, 188 (2019).

\bibitem{Bazeia:2021jok}
D.~Bazeia, D.~A.~Ferreira and M.~A.~Marques,
Eur. Phys. J. C 81, 619 (2021).

\bibitem{Bazeia:2021bwg}
D.~Bazeia and A.~S.~Lob\~ao,
EPL 136, 61002 (2021).

\bibitem{Rosa:2021myu}
J.~L.~Rosa, D.~Bazeia and A.~S.~Lob\~ao,
Eur. Phys. J. C 82, 250 (2022).

\bibitem{Kohri:2022vst}
K.~Kohri and K.~i.~Maeda,
[arXiv:2206.11257 [gr-qc]].

\bibitem{Matsumoto:2022tlr}
A.~Matsumoto, M.~Ouchi, K.~Nakajima, M.~Kawasaki, K.~Murai, K.~Motohara, Y.~Harikane, Y.~Ono, K.~Kushibiki and S.~Koyama, 
\textit{et al.}
[arXiv:2203.09617 [astro-ph.CO]].


\bibitem{Bazeia:2022BL}
D.~Bazeia and A.~S.~Lob\~ao Jr.,
Eur. Phys. J. C 82, 579 (2022).

\bibitem{Bazeia:2006ef}
D.~Bazeia, F.~A.~Brito and L.~Losano,
JHEP 11, 064 (2006).

\bibitem{Ahmed:2012nh}
A.~Ahmed and B.~Grzadkowski,
JHEP 01, 177 (2013).

\bibitem{deSouzaDutra:2014ddw}
A.~de Souza Dutra, G.~P.~de Brito and J.~M.~Hoff da Silva,
Phys. Rev. D 91, 086016 (2015).

\bibitem{Zhong:2017uhn}
Y.~Zhong, Y.~Zhong, Y.~P.~Zhang and Y.~X.~Liu,
Eur. Phys. J. C 78 (2018) no.1, 45.

\bibitem{Zhong:2018fdq}
Y.~Zhong, Y.~P.~Zhang, W.~D.~Guo and Y.~X.~Liu,
JHEP 04 (2019), 154.

\bibitem{Almeida:2009jc}
C.~A.~S.~Almeida, M.~M.~Ferreira, Jr., A.~R.~Gomes and R.~Casana,
Phys. Rev. D 79, 125022 (2009).

\bibitem{Castro:2010uj}
L.~B.~Castro and L.~A.~Meza,
EPL 102, 21001 (2013).

\bibitem{Moreira:2021vcf}
A.~R.~P.~Moreira, J.~E.~G.~Silva and C.~A.~S.~Almeida,
Eur. Phys. J. C 81, 298 (2021).

\bibitem{DWAA}R.M. Hawkins and J.E. Lidsey, Phys. Rev. D 63, 041301 (2001).

\bibitem{DWBB}D. Bazeia, F.A. Brito and F.G. Costa, Phys. Lett. B 661, 179 (2008).

\bibitem{Zhong:2021voa}
Y.~Zhong, F.~Y.~Li and X.~D.~Liu,
Phys. Lett. B 822, 136716 (2021).

\bibitem{Geng:2022slq}
H.~Geng, A.~Karch, C.~Perez-Pardavila, S.~Raju, L.~Randall, M.~Riojas and S.~Shashi, [arXiv:2206.04695 [hep-th]].

\bibitem{Geng:2022tfc}
H.~Geng, [arXiv:2206.11277 [hep-th]].

\bibitem{Rosa:2021tei}
J.~L.~Rosa, M.~A.~Marques, D.~Bazeia and F.~S.~N.~Lobo,
Eur. Phys. J. C 81, 981 (2021).

\bibitem{Rosa:2022fhl}
J.~L.~Rosa, A.~S.~Lob\~ao and D.~Bazeia,
Eur. Phys. J. C 82, 191 (2022).

\bibitem{Bartolo:2021wpt}
N.~Bartolo, A.~Ganz and S.~Matarrese,
JCAP 05, 008 (2022).
\end{thebibliography}
\end{document}